\PassOptionsToPackage{dvipsnames}{xcolor}
\documentclass[sigconf, nonacm]{acmart}

\newcommand\vldbdoi{XX.XX/XXX.XX}
\newcommand\vldbpages{XXX-XXX}
\newcommand\vldbvolume{14}
\newcommand\vldbissue{1}
\newcommand\vldbyear{2020}
\newcommand\vldbauthors{\authors}
\newcommand\vldbtitle{\paperTitle} 
\newcommand\vldbavailabilityurl{}
\newcommand\vldbpagestyle{plain} 

\AtBeginDocument{%
  }

\usepackage[breakable]{tcolorbox}
\tcbuselibrary{skins} %
\usepackage{tabularx} %
\usepackage{xspace}
\PassOptionsToPackage{dvipsnames}{xcolor}  %
\usepackage{xargs}                      %
\usepackage[colorinlistoftodos,prependcaption,textsize=tiny]{todonotes}

\usepackage[ruled,linesnumbered,vlined]{algorithm2e}
\usepackage{enumitem}

\usepackage{multirow}
\usepackage{listings}
\usepackage{colortbl}
\definecolor{lightgray}{gray}{0.9}

\usepackage{cleveref}
\usepackage{fancyvrb}
\usepackage[normalem]{ulem}
\usepackage{setspace}

\usepackage{comment}  %
\usepackage{makecell}
\usepackage{pifont} %

\def\Snospace~{\S{}}

\makeatletter
\newcommand{\crefnames}[3]{%
  \@for\next:=#1\do{%
    \expandafter\crefname\expandafter{\next}{#2}{#3}%
  }%
}
\makeatother
\crefnames{part,chapter,section}{\S}{\S\S}
\crefnames{paragraph,subparagraph}{\P}{\P\P}
\crefname{figure}{Fig.}{Figs.}
\crefname{algorithm}{Algorithm}{Algs.}

\tcbset{querybox/.style={
  colback=gray!5!white, colframe=gray!30,
  boxrule=0.2pt, arc=2pt, left=4pt, right=4pt, top=2pt, bottom=2pt,
  fontupper=\small
}}

\newcommand{\sys}{\mbox{\textsc{Halo}}\xspace}

\newcommand{\paperTitle}{\sys: Domain-Aware Query Optimization for Long-Context Question Answering}

\newcommand{\eg}{\textit{e}.\textit{g}.,\xspace}
\newcommand{\ie}{\textit{i}.\textit{e}.,\xspace}

\newcommand{\PP}[1]{
\vspace{2px}
\noindent{\bf\textsc{#1}.}\xspace
}

\newcommand{\takeaway}[1]{%
\par\vspace{2pt}%
\noindent\textit{\small\textbf{$\Rightarrow$ Takeaway:} #1}%
\vspace{2pt}%
}

\tcbset{
    parasummary/.style={
        colback=yellow!05!white, 
        colframe=black, 
        sharp corners,
        boxrule=0.5mm, 
        width=\linewidth
    }
}

\tcbset{
    interface/.style={
        colback=gray!5,           %
        colframe=black,          %
        coltitle=black,          %
        sharp corners,           %
        boxrule=0.1mm,           %
        fonttitle=\bfseries,     %
        fontupper=\footnotesize,
        fontlower=\footnotesize,
        before skip=1em, %
        after skip=1em, %
        colbacktitle=gray!20,    %
        enhanced,                %
        top=5pt, bottom=5pt,     %
        left=5pt, right=5pt
    }
}

\tcbset{
    feedback/.style={
        colback=red!10!white, 
        colframe=blue!75!black, 
        rounded corners,
        boxrule=0.4mm, 
        width=\linewidth,
        before skip=1em, %
        after skip=1em %
        floatplacement=t!, %
        float
    }
}

\tcbset{
    keytakeaway/.style={
        colback=blue!10!white, 
        colframe=green!75!black, 
        rounded corners,
        boxrule=0.4mm, 
        width=\linewidth,
        before skip=1em, %
        after skip=1em %
        floatplacement=t!, %
        float
    }
}

\tcbset{
    thingstoadd/.style={
        colback=red!10!white, 
        colframe=black!50, 
        enhanced jigsaw, %
        boxrule=0.3mm, 
        width=\linewidth
    }
}

\newcommandx{\unsure}[2][1=]{\todo[linecolor=red,backgroundcolor=red!25,bordercolor=red,#1]{#2}}
\newcommandx{\change}[2][1=]{\todo[linecolor=blue,backgroundcolor=blue!25,bordercolor=blue,#1]{#2}}
\newcommandx{\info}[2][1=]{\todo[linecolor=OliveGreen,backgroundcolor=OliveGreen!25,bordercolor=OliveGreen,#1]{#2}}
\newcommandx{\improvement}[2][1=]{\todo[linecolor=Plum,backgroundcolor=Plum!25,bordercolor=Plum,#1]{#2}}
\newcommandx{\thiswillnotshow}[2][1=]{\todo[disable,#1]{#2}}

\newcommand{\squishitemize}{
 \begin{list}{$\bullet$}
  { \setlength{\itemsep}{0pt}
     \setlength{\parsep}{0pt}
     \setlength{\topsep}{0pt}
     \setlength{\partopsep}{0pt}
     \setlength{\leftmargin}{1.95em}
     \setlength{\labelwidth}{1.5em}
     \setlength{\labelsep}{0.5em} } }

\newcounter{Lcount}
\newcommand{\squishlist}{
    \begin{list}{\arabic{Lcount}. }
   { \usecounter{Lcount}
        \setlength{\itemsep}{0pt}
        \setlength{\parsep}{3pt}
        \setlength{\topsep}{0pt}
        \setlength{\partopsep}{0pt}
        \setlength{\leftmargin}{2em}
        \setlength{\labelwidth}{1.5em}
        \setlength{\labelsep}{0.5em} } }

\newcommand{\squishend}{\end{list}}

\newcommand{\singlepass}{\mbox{\textsc{VanillaLLM}}\xspace}
\newcommand{\singlepassdk}{\mbox{\textsc{VanillaLLM\,+\,DK}}\xspace}
\newcommand{\lcrag}{\mbox{\textsc{RAG}}\xspace}
\newcommand{\lcragdk}{\mbox{\textsc{RAG\,+\,DK}}\xspace}

\newcommand{\multiagent}{\mbox{\textsc{CoA}}\xspace}

\newcommand{\halonodk}{\mbox{\textsc{\sys (No DK)}}\xspace}

\definecolor{webblue}{RGB}{0, 76, 153} %

\newcommand{\parser}{\mbox{\textsc{KnowledgeParser}}\xspace}
\newcommand{\selector}{\mbox{\textsc{ContextSelector}}\xspace}
\newcommand{\inferenceengine}{\mbox{\textsc{InferenceEngine}}\xspace}
\newcommand{\verifier}{\mbox{\textsc{Verifier}}\xspace}

\newcommand{\segment}{\mbox{\textsc{Structural}}\xspace}
\newcommand{\filter}{\mbox{\textsc{Filter}}\xspace}
\newcommand{\validate}{\mbox{\textsc{Validate}}\xspace}

\definecolor{bg}{rgb}{0.95,0.95,0.95}
\definecolor{keywordcolor}{rgb}{0.6,0.1,0.8}
\definecolor{stringcolor}{rgb}{0.0,0.5,0.5}
\definecolor{commentcolor}{rgb}{0.4,0.4,0.4}

\lstdefinelanguage{HintLang}{
    morekeywords={Segment,Filter,Validate},
    sensitive=true,
    morestring=[b]",
    alsoletter={=},
}

\lstdefinestyle{hintstyle}{
    language=HintLang,
    backgroundcolor=\color{bg},
    basicstyle=\ttfamily\footnotesize,
    keywordstyle=\color{keywordcolor}\bfseries,
    stringstyle=\color{stringcolor},
    showstringspaces=false,
    breaklines=true,
    frame=single,
    tabsize=2,
    columns=fullflexible
}

\begin{document}

\title{\paperTitle}

\author{Pramod Chunduri}
\authornote{Work done prior to joining Amazon Web Services.}
\affiliation{%
  \institution{Amazon Web Services}
    \country{}}
\email{cnpramod@amazon.com}

\author{Francisco Romero}
\affiliation{%
  \institution{Georgia Institute of Technology}
    \country{}}
\email{faromero@gatech.edu}

\author{Ali Payani}
\affiliation{%
  \institution{Cisco}
    \country{}}
\email{apayani@cisco.com}

\author{Kexin Rong}
\affiliation{%
  \institution{Georgia Institute of Technology}
    \country{}}
\email{krong@gatech.edu}

\author{Joy Arulraj}
\affiliation{%
  \institution{Georgia Institute of Technology}
    \country{}}
\email{jarulraj@gatech.edu}

\begin{abstract}

Long-context question answering (QA) over lengthy documents is critical for applications such as financial analysis, legal review, and scientific research.
Current approaches, such as processing entire documents via a single LLM call or retrieving relevant chunks via RAG have two drawbacks:
First, as context size increases, response quality can degrade, impacting accuracy.
Second, iteratively processing hundreds of input documents can incur prohibitively high costs in API calls.
To improve response quality and reduce the number of iterations needed to get the desired response, users tend to add domain knowledge to their prompts.
However, existing systems fail to systematically capture and use this knowledge to guide query processing.
Domain knowledge is treated as prompt tokens alongside the document: the LLM may or may not follow it, there is no reduction in computational cost, and when outputs are incorrect, users must manually iterate.

We present Halo, a long-context QA framework that automatically extracts domain knowledge from user prompts and applies it as executable operators across a multi-stage query execution pipeline.
Halo identifies three common forms of domain knowledge -- where in the document to look, what content to ignore, and how to verify the answer -- and applies each at the pipeline stage where it is most effective: pruning the document before chunk selection, filtering irrelevant
chunks before inference, and ranking candidate responses after generation.
To handle imprecise or invalid domain knowledge, Halo uses a fallback mechanism that detects low-quality operators at runtime and selectively disables them.
Our evaluation across finance, literature, and scientific datasets shows that Halo achieves up to 13\% higher accuracy and 4.8$\times$ lower cost compared to baselines, and enables a lightweight open-source model to approach frontier LLM accuracy at 78$\times$ lower cost.
\end{abstract}

\maketitle

\pagestyle{\vldbpagestyle}
\begingroup\small\noindent\raggedright\textbf{PVLDB Reference Format:}\\
\vldbauthors. \vldbtitle. PVLDB, \vldbvolume(\vldbissue): \vldbpages, \vldbyear.\\
\href{https://doi.org/\vldbdoi}{doi:\vldbdoi}
\endgroup
\begingroup
\renewcommand\thefootnote{}\footnote{\noindent
This work is licensed under the Creative Commons BY-NC-ND 4.0 International License. Visit \url{https://creativecommons.org/licenses/by-nc-nd/4.0/} to view a copy of this license. For any use beyond those covered by this license, obtain permission by emailing \href{mailto:info@vldb.org}{info@vldb.org}. Copyright is held by the owner/author(s). Publication rights licensed to the VLDB Endowment. \\
\raggedright Proceedings of the VLDB Endowment, Vol. \vldbvolume, No. \vldbissue\ %
ISSN 2150-8097. \\
\href{https://doi.org/\vldbdoi}{doi:\vldbdoi} \\
}\addtocounter{footnote}{-1}\endgroup

\ifdefempty{\vldbavailabilityurl}{}{
\vspace{.3cm}
\begingroup\small\noindent\raggedright\textbf{PVLDB Artifact Availability:}\\
The source code, data, and/or other artifacts have been made available at \url{\vldbavailabilityurl}.
\endgroup
}

\section{Introduction}
\label{sec:introduction}
Long-Context Question Answering (QA), an important task in document understanding, aims to answer fact-based questions over lengthy documents.
Real-world applications, such as financial analysts querying 10-K filings, legal professionals reviewing contracts, and researchers surveying scientific literature, regularly demand accurate answers from documents spanning hundreds of thousands of tokens~\cite{docfinqa, wangnovelqa}.
Recent advances in large language models (LLMs) have significantly expanded their \textit{context windows}, with some models such as Gemini 3 Pro and Sonnet 4.5 supporting a million tokens~\cite{googleGeminiAPI2026, sonnet452025}.
However, efficiently leveraging these contexts for QA remains challenging.

Currently, there are two standard approaches for tackling long-context QA, yet neither achieves a satisfactory accuracy-cost tradeoff.
The first feeds the entire document to a state-of-the-art (SoTA) LLM like Sonnet 4.5 in a single call, which we refer to as \singlepass.
However, this approach is both expensive and unreliable: processing hundreds of long documents can cost over \$500 in API calls (\autoref{sec:eval:end-to-end})
and even SoTA LLMs exhibit significant hallucination over long contexts due to issues like \textit{``lost in the middle''}~\cite{lostinthemiddle, liu2025towards}.
The second approach is Retrieval-Augmented Generation (RAG), which attempts to reduce cost by selecting only the most relevant document chunks before inference~\cite{ragoriginal}.
Although RAG reduces token usage compared to \singlepass, it introduces top-K selection errors that could hurt accuracy: key evidence may be missed during the similarity search, and fragmented or reordered chunks can introduce incoherent context.
As a result, RAG still hallucinates or overlooks important nuances in the provided context~\cite{leng2024long, longcontextrag}.

\begin{figure}[t]
    \centering    \includegraphics[width=\columnwidth]{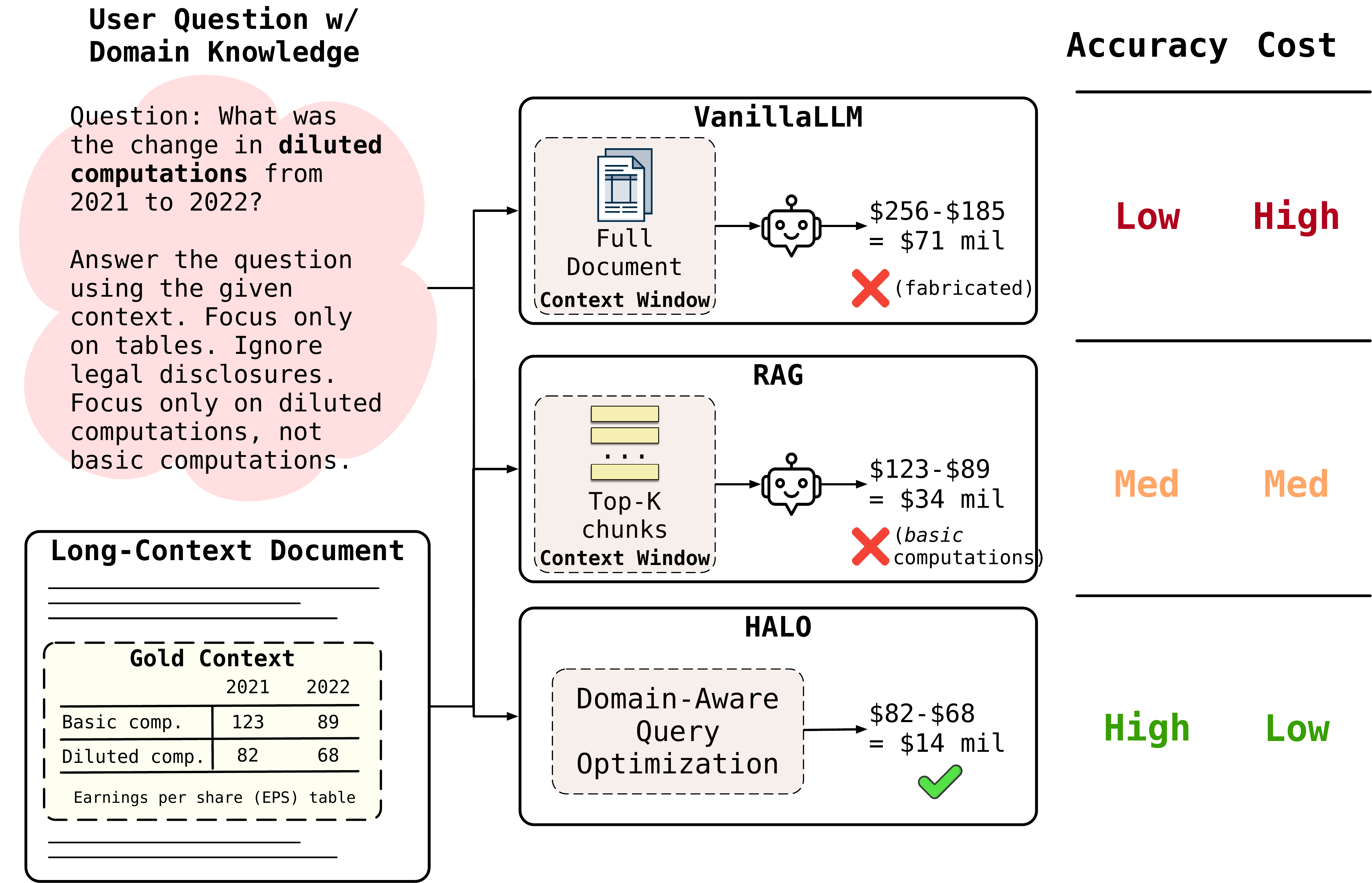}
    \caption{\singlepass and \lcrag cannot utilize user's domain knowledge effectively for long-context QA. \sys extracts this knowledge and systematically applies it in a multi-stage query optimization pipeline.}
    \vspace{-2mm}
    \label{fig:baselines-example}
\end{figure}

To improve the accuracy-cost tradeoff, domain experts naturally turn to domain knowledge.
However, the primary mechanism available to them is prompt engineering, where such knowledge is embedded directly into the model's prompt input, as shown in \autoref{fig:baselines-example}.
For example, a financial analyst may instruct the LLM to focus only on tables or ignore boilerplate disclosures when processing SEC 10-K filings.
While such instructions can sometimes improve answer quality, they remain embedded as free-form text within the prompt.
The model may inconsistently follow the instructions, especially in long contexts where instruction-following degrades~\cite{he2024can,yen2025helmet}. 
As seen in~\autoref{fig:baselines-example}, \singlepass fabricates 
incorrect values even when domain knowledge is provided in the 
prompt as the instructions are lost in the 200K-token context.
\lcrag retrieves relevant context but hallucinates the final response, reporting basic instead of diluted computations, due to the incoherent top-K chunks.
In both cases, the domain knowledge yields no cost reduction.
Moreover, when such errors occur, users must iteratively refine prompts through trial and error.
The core limitation is that domain knowledge is expressed implicitly as tokens rather than captured explicitly as structured guidance that shapes how the system processes the query, limiting improvements in both accuracy and cost.

\begin{figure*}[t]
    \centering
    \includegraphics[width=0.9\linewidth]{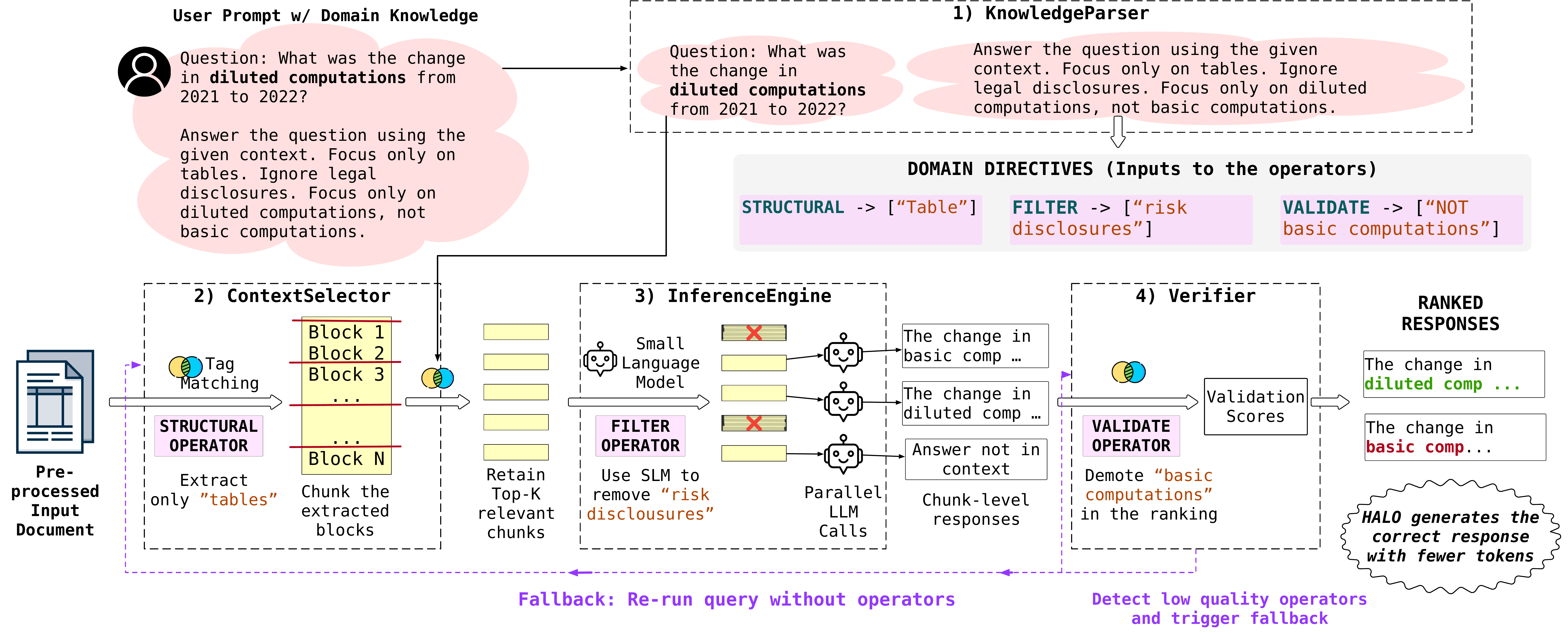}
    \caption{Workflow of \sys: (1) the \parser extracts domain directives from the user prompt, (2) the \selector prunes structural elements using the \segment operator and identifies the top-$k$ relevant chunks, (3) the \inferenceengine applies the \filter operator to discard irrelevant chunks and generates candidate responses on the remaining chunks, (4) the \verifier re-ranks the responses using \validate operator--demoting hallucinatory responses.}
    \label{fig:running-example}
\end{figure*}

\PP{Our Operator-Based Approach}
To address this limitation, we introduce~\sys, a long-context QA framework that extracts domain knowledge from free-form prompt text into explicit \textit{directives} that directly shape execution plans.
Our key observation is that domain knowledge should not be consumed as prompt tokens for the LLM to implicitly interpret and instead be extracted into structured directives that are applied at the pipeline stage where each is most effective.
\sys identifies three common forms of domain knowledge from users' natural language prompts:
\squishitemize
    \item \segment---\textit{where to look}: structural elements likely to contain the answer (\eg tables, MD\&A sections)
    \item \filter---\textit{what to ignore}: semantic content irrelevant to the query (\eg legal disclosures, boilerplate text)
    \item \validate---\textit{how to verify}: domain-specific constraints to demote hallucinated responses (\eg ``NOT basic computations'')
\squishend
The \parser first extracts structured directives from the user's unstructured prompt.
\sys then leverages \textit{operators} to apply these directives across a multi-stage pipeline (\autoref{fig:running-example}), with each operator designed around a specific insight about where domain knowledge has the highest impact.
The \selector stage applies the \segment operator to prune the document to only the user-specified structural elements \textit{before} chunking.
By carefully curating the expensive LLM’s context, \sys ensures only relevant context is passed downstream to the other operators.
The \inferenceengine stage then applies the \filter operator via a lightweight SLM to discard irrelevant chunks \textit{before} the expensive LLM processes them.
It uses a model cascade where a cheap model quickly eliminates noise so the expensive model only sees what matters.
Finally, the \verifier stage applies the \validate operator to rank candidate responses.
We observe that positive and negative validation directives (\eg ``diluted EPS'' vs.\ ``basic EPS'') can be contrasted against each other: responses aligned with positive directives are promoted while those matching negative directives are demoted, enabling domain-aware hallucination detection without any additional LLM calls.
To prevent accuracy degradation from low-quality operators, \sys's fallback manager detects when an operator hurts performance using model- and operator-based signals, and disables it to recover accuracy at minimal overhead.

Our evaluation across finance, literature, and scientific datasets shows that \sys achieves up to 13\% higher accuracy and 4.8$\times$ lower cost compared to baselines.
Notably, \sys enables a lightweight open-source model to approach the accuracy of a frontier LLM at 78$\times$ lower cost, while the domain operators themselves add only 2\% overhead to the total query cost.

\PP{Contributions} We make the following key contributions:
\squishitemize
\item We show the domain knowledge users express in 
  their prompts can be automatically extracted into structured 
  directives and design corresponding operators (\segment, \filter, \validate) that each apply the directives
  (\autoref{sec:capturing-domain}, \autoref{sec:applying-domain}).
\item We propose \sys, a multi-stage long-context QA 
  framework that composes these operators to simultaneously 
  reduce cost and improve accuracy--pruning document 
  structure early, filtering chunks via lightweight models, and ranking responses using domain 
  constraints (\autoref{sec:overview}, 
  \autoref{sec:applying-domain}).
\item We develop a fallback mechanism that detects when 
  operators degrade accuracy at runtime and selectively 
  disables them, ensuring \sys effectively handles low quality directives (\autoref{sec:fallback-manager}).
\item We evaluate \sys across diverse domains and datasets,
  showing up to 13\% higher accuracy 
  and 4.8$\times$ lower cost over baselines, and 
  demonstrating that a lightweight open-source model with 
  \sys approaches frontier LLM accuracy at 78$\times$ lower 
  cost (\autoref{sec:evaluation}).
\squishend

\section{Motivating Example}
\label{sec:motivating-example}

To illustrate the value of domain knowledge for long-context QA and the challenges of applying it systematically, consider the following example.
A junior financial analyst is tasked with extracting specific numerical data from hundreds of annual SEC 10-K filings.
Each filing is roughly 200K tokens, containing financial tables, narrative discussions, legal disclosures, and boilerplate text.
The analyst wishes to answer questions like:

\smallskip
\noindent \textit{``What was the change in diluted computations from 2021 to 2022?''}
\smallskip

\noindent Answering this question requires locating the correct earnings-per-share table, extracting the ``diluted'' (not ``basic'') computations for the correct years, and computing the difference.
The analyst must do this across hundreds of documents, making manual inspection infeasible at scale.

\PP{Baseline Approaches}
We first consider two standard approaches that do not leverage domain knowledge.

\noindent \textit{Vanilla LLM.}
A state-of-the-art long-context LLM such as Claude Sonnet 4.5 can process the entire 200K-token filing in a single call.
However, this approach is both expensive and unreliable.
Processing hundreds of documents can cost over \$500 in API calls (\autoref{sec:eval:end-to-end}).
Furthermore, the LLM is prone to hallucinate over such long contexts~\cite{lostinthemiddle, liu2025towards}.
For the query above, a SoTA LLM fabricates the values for 2021-2022, simply because it could not find the relevant context from the long document.

\noindent \textit{RAG.}
To avoid processing the entire document, a standard RAG pipeline~\cite{ragoriginal} retrieves the most relevant chunks and provides them as context to the LLM.
While this reduces cost, it also introduces retrieval errors: for the query above, RAG ranks a chunk containing \textit{basic} computations higher than the one with \textit{diluted} computations, and the LLM confidently answers based on the wrong data.

\noindent In both approaches, the analyst's domain knowledge plays no role in the processing pipeline.
Furthermore, when outputs are incorrect, both approaches require iterative prompt refinement---compounding cost and effort across hundreds of documents.

\PP{Benefits of Domain Knowledge}
To improve the LLM's response quality, analysts can use domain knowledge to optimize their queries:

\squishitemize
\item \textit{Structural knowledge: where to look.}
The analyst knows that earnings-per-share data is always in a table, never in narrative text.
Pruning non-tabular content reduces the search space from 200K to 50K tokens.

\item \textit{Filtering knowledge: what to ignore.}
The analyst also knows that large portions of the filing, such as legal disclosures, risk factors, and forward-looking statements, are never relevant to earnings queries.
For example, removing boilerplate legal tables from the 50K tokens of tabular content could reduce the context to approximately 30K tokens, eliminating noise that might otherwise confuse the model.

\item \textit{Validation knowledge: how to verify.}
The analyst knows that the earnings-per-share table contains both ``basic'' and ``diluted'' computations, and the LLM frequently confuses them.
If the processing pipeline could use this knowledge to verify that a response references ``diluted'' rather than ``basic'' computations, accuracy could be improved.
\squishend

As shown in~\autoref{tab:benefits-of-domain-knowledge}, each type of domain knowledge provides a distinct benefit.
Structural knowledge narrows the search space by pruning irrelevant document regions, improving the quality of retrieved chunks.
Filtering knowledge further reduces cost by discarding irrelevant chunks before passing them to the expensive LLM.
Validation knowledge improves accuracy by detecting and mitigating hallucinations.

\begin{table}[t]
\centering
\caption{Potential benefits of domain knowledge for the example financial QA query. Each row adds one type of knowledge on top of the previous.}
\vspace{-2mm}
\label{tab:benefits-of-domain-knowledge}
\resizebox{\columnwidth}{!}{%
\begin{tabular}{lccc}
\toprule
\textbf{Approach} & \textbf{Tokens} & \textbf{Result} & \textbf{Knowledge Used} \\
\midrule
Vanilla LLM                     & $\sim$200K & \ding{55} (wrong years)            & None \\
RAG (top-K)                     & $\sim$100K  & \ding{55} (basic, not diluted)     & None \\
\midrule
+ Structural knowledge          & $\sim$50K  & \ding{55} (basic, not diluted)     & Where to look \\
+ Filtering knowledge           & $\sim$30K  & \ding{55} (basic, not diluted)     & + What to ignore \\
+ Validation knowledge          & $\sim$30K  & \ding{51} (correct)                & + How to verify \\
\bottomrule
\end{tabular}%
}
\end{table}

\PP{Applying Domain Knowledge Today}
Despite its clear value, applying domain knowledge to long-context QA today is a manual, tedious, and unreliable process.

\noindent \textit{Prompt engineering.}
The most common approach is to embed domain knowledge directly into the LLM prompt (\autoref{fig:baselines-example}).
The LLM receives these instructions as input tokens alongside the document.
It may or may not follow them: the model might focus on tables but still processes the entire document, offering no cost reduction.
Whether the model ignores legal disclosures or avoids ``basic computations'' is unknown to the analyst until they inspect the output.

\definecolor{lotusop}{RGB}{180, 40, 90}
\begin{figure}[t]
\centering
\includegraphics[width=\columnwidth]{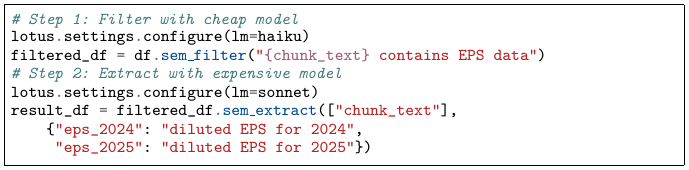}
\caption{LOTUS program for extracting EPS data from financial filings. The user manually specifies the filter predicate and extraction schema in Python.}
\label{fig:lotus-example}
\end{figure}

\noindent \textit{Manual iteration.}
When the output is incorrect, the analyst must inspect the response, diagnose the error, refine the prompt, and re-run the query.
This iteration loop is expensive: each attempt processes the full document, and the analyst may need multiple rounds to converge on a working prompt, costing 3-5$\times$ more than the already expensive \singlepass.

\noindent \textit{Manual pipeline management.}
To realize the cost benefits shown in~\autoref{tab:benefits-of-domain-knowledge}, the analyst would need to go beyond prompt engineering and manually build a processing pipeline.
Systems like LOTUS~\cite{lotus} provide semantic operators for LLM-based data processing, but require users to build LLM pipelines that manually specify filter predicates and extraction schemas (\autoref{fig:lotus-example}).
For our financial QA example, the analyst must determine the correct filter keywords, define the output schema, specify the models to use, and perform the final computation manually---a cumbersome burden for domain experts who do not specialize in building LLM pipelines.

\PP{Key observations}
From the financial analyst example, we identify two key takeaways:
(1) domain knowledge is valuable for long-context QA, providing improvements in both cost and accuracy if used correctly, and
(2) applying this knowledge today is manual and expensive.
To address this, we need to \emph{automatically capture the domain knowledge users already express in their prompts}, and \emph{systematically apply it during query optimization and processing}.

\section{System Overview} 
\label{sec:overview}

We now describe \sys: a framework for long-context question answering that automatically extracts domain knowledge from user prompts and systematically applies it to optimize query processing.
Unlike existing approaches that treat domain knowledge as prompt tokens for the LLM to implicitly interpret, \sys extracts this knowledge into structured \textit{directives} and applies them via dedicated \textit{operators} across a multi-stage pipeline.
\autoref{fig:running-example} illustrates the end-to-end workflow.

\subsection{End-to-end workflow}
\sys takes the same input a user would provide to any LLM: a document and a prompt containing the question along with domain knowledge for guiding the response.
Our experiments across finance, literature, and scientific domains (\autoref{sec:evaluation}) identify three common forms of domain knowledge:
(1) \textit{structural} knowledge of where in the document the answer is likely found (\eg a table),
(2) \textit{filtering} knowledge of what content to ignore (\eg legal disclosures), and
(3) \textit{validation} knowledge for detecting hallucinations (\eg confusable values).
\sys extracts these into three directive types---$\mathcal{S}$, $\mathcal{F}$, and $\mathcal{V}$---and applies them via the \segment, \filter, and \validate operators, respectively (\autoref{sec:capturing-domain}, \autoref{sec:applying-domain}).
The framework is extensible to additional operator types (\autoref{sec:applying-domain:extensions}).

As shown in~\autoref{fig:running-example}, \sys consists of five stages: a one-time preprocessing stage and four query-time stages:

\begin{figure}[t]
    \centering
    \includegraphics[width=0.7\columnwidth]{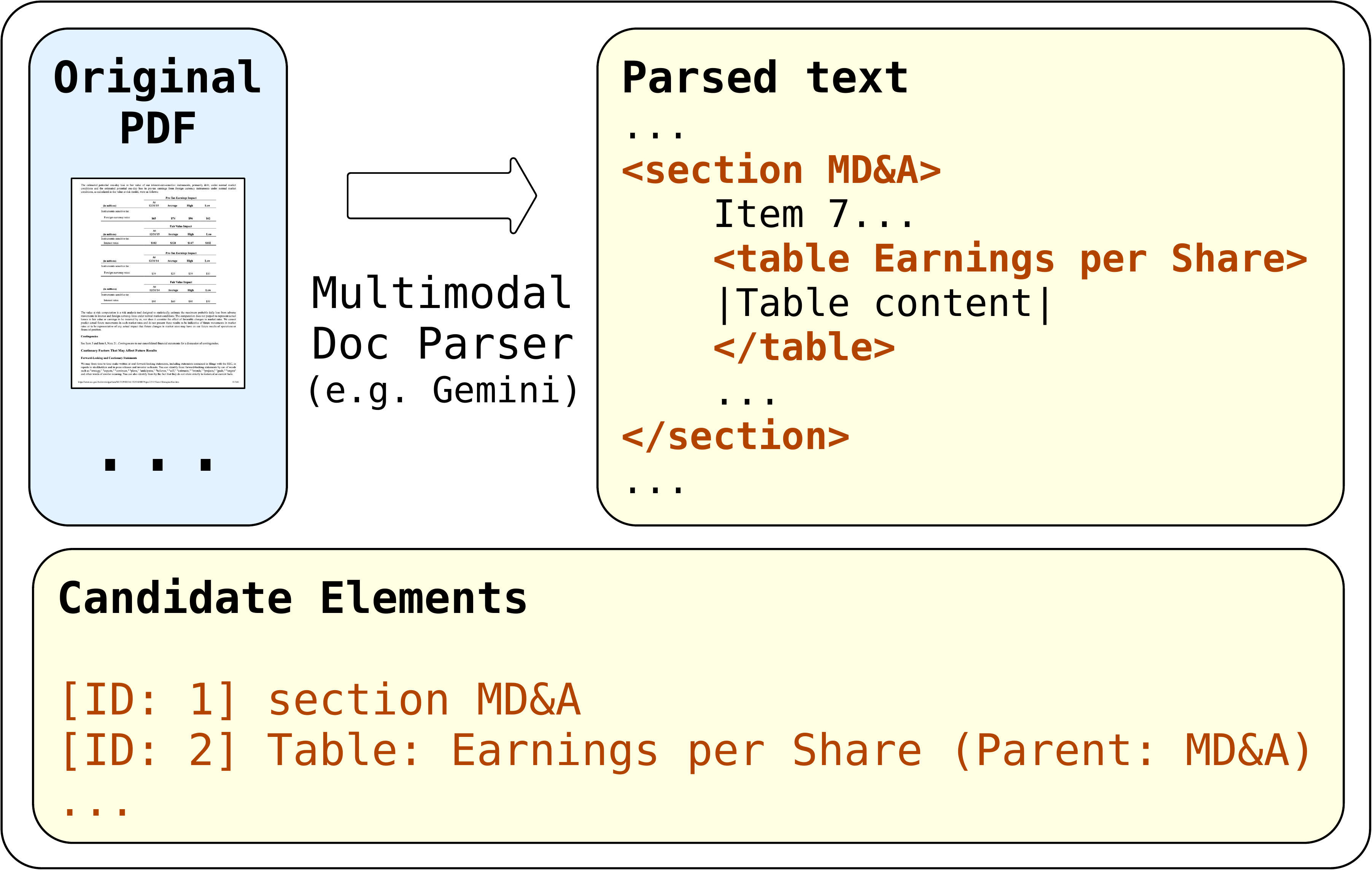}
    \vspace{-0.5em}
    \caption{Structure-aware document preprocessing}
    \vspace{-0.5em}
    \label{fig:preprocessing}
\end{figure}
\noindent{\textbf{(0) Input Preprocessing.}}
This is a one-time stage that parses the PDF document into a structure-aware representation $\tilde{D}$ using a multimodal parsing utility~\cite{filimonov2025gemini}, as shown in~\autoref{fig:preprocessing}.
The extracted structural elements (\eg tables, sections, figures) are available for all subsequent queries over the document.
While LLM-based PDF parsing is expensive, it is highly accurate~\cite{filimonov2025gemini} and the cost is amortized across multiple queries (~\autoref{sec:eval-overhead}).

\noindent{\textbf{(1) \parser.}}
This stage uses an LLM to extract domain directives from the user's prompt: $\pi(d) \rightarrow (\mathcal{S}, \mathcal{F}, \mathcal{V})$.
This step processes only the instruction portion of the prompt, incurring negligible cost compared to the full document (\autoref{sec:eval-overhead}).
The extraction is analogous to text-to-SQL parsing with a constrained directive taxonomy~\cite{gan2021natural}.
Once extracted, the directives are consumed by operators across the remaining pipeline stages.

\noindent{\textbf{(2) \selector.}}
This stage applies the \segment operator using the structural directives $\mathcal{S}$ to prune $\tilde{D}$ to only the user-specified structural elements (\eg tables).
This produces a pruned document, $D_s$.
$D_s$ is split into chunks and ranked by embedding similarity to the query; the top-$k$ are selected for downstream processing.
By narrowing the search space, \sys improves the relevance of selected chunks, leading to higher-quality context for downstream inference (\autoref{sec:applying-domain:optimizing-context-sel}).

\noindent{\textbf{(3) \inferenceengine.}}
This stage applies the \filter operator using the filter directives $\mathcal{F}$ to prune irrelevant chunks from the top-$k$ candidates via a lightweight SLM.
The retained chunks are then processed in parallel by the expensive LLM to generate chunk-level candidate responses (\autoref{sec:applying-domain:optimizing-inference}).

\noindent{\textbf{(4) \verifier.}}
This stage applies the \validate operator using the validation directives $\mathcal{V}$ to rank candidate responses.
Each response is scored by its alignment with positive constraints $\mathcal{V}^+$ (what the answer should contain) and negative constraints $\mathcal{V}^-$ (what it should not).
The top-$k$ ranked responses are returned to the user (\autoref{sec:applying-domain:mitigating-hallucinations}).

To handle imprecise directives, \sys includes a fallback manager that detects when operators degrade accuracy---using signals such as universal answer refusal (for \segment/\filter) or low score variance (for \validate)---and selectively disables problematic operators to recover baseline performance (\autoref{sec:fallback-manager}).

\subsection{Example Workflow}
We illustrate \sys's workflow using the financial analyst example from~\autoref{sec:motivating-example}.
The analyst queries a 200K-token SEC 10-K filing: ``What was the change in \textit{diluted computations} from 2021 to 2022?''
The prompt includes domain-specific instructions such as ``focus on tables,'' ``ignore legal disclaimers,'' and ``report diluted computations, NOT basic.'' (\autoref{fig:running-example}).
The \parser extracts these into directives:
$\sigma_1 = \text{``table''}$, \;
$\phi_1 = \text{``legal disclaimers''}$, \;
$\delta_1 = \text{``diluted computations''}$, \;
$\delta_2 = \text{``NOT basic computations''}$.

The \selector applies the \segment operator with $\mathcal{S} = \{\sigma_1\}$ to extract only tabular content, reducing the document from 200K to approximately 50K tokens.
The \inferenceengine applies the \filter operator with $\mathcal{F} = \{\phi_1\}$ to remove boilerplate legal tables, further reducing the candidate set.
The retained chunks are processed via parallel LLM calls to generate candidate responses.
Finally, the \verifier applies the \validate operator with $\mathcal{V}^+ = \{\delta_1\}$ and $\mathcal{V}^- = \{\delta_2\}$ to rank responses, penalizing those mentioning ``basic computations'' and prioritizing those with ``diluted computations.''
This workflow reduces the tokens processed by the expensive LLM while improving accuracy by ensuring the response satisfies domain-specific constraints.

\section{Extracting Domain Knowledge}
\label{sec:capturing-domain}

Today, users often express domain knowledge as plain-text instructions embedded in prompts for LLM-based question-answer workflows.
These instructions are treated as ordinary input tokens—they do not deterministically affect query planning or execution.

\sys instead parses natural-language domain instructions into structured \textit{directives} that directly influence query optimization and execution.
Given a question $q$ and domain knowledge $d$, \sys uses a \parser to extract $\pi(d) \rightarrow (\mathcal{S}, \mathcal{F}, \mathcal{V})$
where:

\squishitemize
    \item $\mathcal{S} = \{\sigma_1, \ldots, \sigma_n\}$ is a set of structural directives
    \item $\mathcal{F} = \{\phi_1, \ldots, \phi_m\}$ is a set of filter directives
    \item $\mathcal{V} = \{\delta_1, \ldots, \delta_k\}$ is a set of validation directives
\squishend

Any of these sets may be empty if the corresponding knowledge is not present in $d$.
Each directive is a declarative specification--it describes \textit{what} knowledge the user expressed, not \textit{how} it is enforced.
Operators, described in Section~\ref{sec:applying-domain}, implement \emph{how} these directives are applied at different stages of execution.
We now describe each domain knowledge type the \parser extracts.

\subsection{Where to focus - Structural directives ($\mathcal{S}$)}
\label{sec:capturing:struct}
Domain experts possess insights about the document structure and where to look in the document when answering questions.
For example, a financial analyst reviewing company earnings may know the earnings per share is located in a table of the SEC 10-K filing.
However, they may not know the exact table.
A scientific researcher who is comparing multiple baselines knows to survey the evaluation section of the paper.

Through their prompts, users provide structural knowledge that may include: (1) structural element types (\eg tables, text, graphs, and figures), and (2) section and table names (\eg ``MD\&A section,'' ``Evaluation section,'' ``EPS table'').
The \parser extracts a set of structural elements $\mathcal{S} = (\sigma_1, \ldots, \sigma_n)$ from $d$, where each $\sigma_i$ is an element from the pre-determined document structure.
For example, from the instruction ``the answer is likely in a table,'' the \parser extracts $\sigma_1 = \text{``table''}$.

\sys supports structural knowledge of arbitrary granularity, such as `look in the MD\&A section' or `the answer is in the EPS table row 4'.
The extracted structural elements and how they are utilized by \sys to optimize query execution are discussed in~\autoref{sec:applying-domain:optimizing-context-sel}.

\subsection{What to ignore - Filter directives ($\mathcal{F}$)}
\label{sec:capturing:filter}

Prompts often include domain knowledge about where not to look or what to skip.
For example, a financial analyst may provide instructions to skip over boilerplate content like `risk disclosures' in financial documents.
A paralegal may prompt the LLM to ignore `waivers' in service-level agreements (SLA).

While structural directives focus on document structure, filter directives focus on \textit{document content}---``risk disclosures'' and ``waivers'' are semantic concepts that span arbitrary structural elements.
For instance, ``risk disclosures'' appear in multiple sections of the SEC 10-K document, containing multiple structural types such as tables, text paragraphs, and footnotes.

The \parser extracts filter directives $\mathcal{F} = (\phi_1, \ldots, \phi_n)$ from $d$, where each $\phi_i$ is a natural language description of the content to exclude.
For example, for the instruction ``ignore general risk disclosures and legal disclaimers that are not relevant to the current query,'' the filter criteria are extracted as:
\begin{align*}
\phi_1 = \text{``risk disclosures''}, \phi_2 = \text{``legal disclaimers''}
\end{align*}

These extracted filter directives are applied during \sys's query execution phase to exclude matching content (discussed in~\cref{sec:applying-domain:optimizing-inference}).

\subsection{How to verify - Validation directives ($\mathcal{V}$)}
\label{sec:capturing:validate}
To avoid hallucinations, users commonly augment their prompts with instructions on how to disambiguate requested information from extraneous content and what errors to avoid when answering the question.
Users may provide disambiguation knowledge as:
\squishitemize
\item Emphasis constraints: ``only return numerical values,'' ``do not include commentary''
\item Disambiguation: ``diluted EPS, NOT basic EPS,'' ``net revenue, NOT gross''
\squishend

The \parser extracts such knowledge into a set of validation directives $\mathcal{V} = (\delta_1, \ldots, \delta_n)$ where each $\delta_i$ is an emphasis constraint or disambiguation criterion.
For example, the instruction ``return diluted EPS only, NOT basic EPS'' is extracted into:
\begin{align*}
\delta_1 = \text{``diluted EPS''}, \delta_2 = \text{``NOT basic EPS''}
\end{align*}
The \parser identifies negative constraints syntactically (\eg via ``NOT'', 
``avoid'', ``do not include'') and partitions $\mathcal{V}$ into positive 
directives $\mathcal{V}^+$ (what the response \textit{should} contain) and 
negative directives $\mathcal{V}^-$ (what it should \textit{not} contain).
For example, $\delta_1 = \text{``diluted EPS''} \in \mathcal{V}^+$ and 
$\delta_2 = \text{``basic EPS''} \in \mathcal{V}^-$.
We discuss how the validation directives are used to detect and mitigate model hallucinations in~\cref{sec:applying-domain:mitigating-hallucinations}.

\section{Applying Domain Knowledge}
\label{sec:applying-domain}
Once the \parser extracts the domain knowledge into the directives, these directives must be applied to the query execution pipeline to improve query processing efficiency and accuracy.
In this section, we describe how the directives discussed in~\autoref{sec:capturing-domain} are applied to enable long-context query optimization.
\subsection{Directive-driven Query Execution}
\label{sec:applying-domain:directive-driven}
Existing QA frameworks treat domain instructions $d$ as part of the model input.
In a vanilla LLM pipeline, 
\[ LLM(q, D, d) \rightarrow a, \]
where the query $q$, input document $D$ and natural-language domain instructions $d$ are concatenated into a single prompt.

In retrieval-augmented generation (RAG), the input $D$ is split into chunks $\mathcal{C} = \{c_1, c_2, \dots, c_n\}$.
At query time, a retrieval function $R$ ranks chunks by embedding similarity and selects the top-$k$:
\[ R(q, \mathcal{C}, k) \rightarrow \mathcal{C}_k \quad \text{where} \quad \mathcal{C}_k \subseteq \mathcal{C}, |\mathcal{C}_k| = k \]
The retrieved chunks are concatenated and sent to the model:
\[ RAG(q, \mathcal{C}_k, d) \rightarrow a \]

\noindent \textbf{Key Insight.}
In both approaches, domain instructions $d$ are treated as part of the prompt --- they are not explicitly used to curate context or detect hallucinations.
Users embed valuable domain knowledge in $d$ (\eg ``look in tables,'' ``ignore legal disclaimers''), but this knowledge is not leveraged for query optimization.

\noindent \textbf{\sys Execution.}
\sys extracts domain directives from $d$ using the \parser: $\pi(d) \rightarrow (\mathcal{S}, \mathcal{F}, \mathcal{V})$ and applies them through dedicated operators during execution.
\sys introduces three operators—\segment (structural pruning), \filter (chunk filtering), and \validate (response ranking)—each designed to apply a distinct class of directives at the appropriate stage of the pipeline.
\autoref{tab:knowledge-directives-operators} summarizes the mapping from domain knowledge to directives and their enforcing operators.
\sys's full execution pipeline is as follows:
\vspace{-0.8em}
\begin{align*}
D_s           &= \segment(\tilde{D},\; \mathcal{S})              \notag\\
\mathcal{C}_k &= \text{Sel}(D_s,\; q)                            \notag\\
\mathcal{C}^* &= \filter(\mathcal{C}_k,\; \mathcal{F},\; q)      \notag\\
\mathcal{R}   &= \text{Gen}(\mathcal{C}^*,\; q)                  \notag\\
\sys(q, D, d) &= \validate(\mathcal{R},\; \mathcal{V})
\label{eq:pipeline}
\end{align*}
where $\tilde{D}$ is structure-aware representation of the input document generated in the preprocessing stage (\cref{sec:overview}), $\text{Sel}$ ranks chunks of $D_s$ by embedding similarity to $q$ and selects the top-$k$, 
and $\text{Gen}$ produces a candidate response per chunk via separate LLM calls.
In the subsequent sections, we describe each operator and its impact on the query execution cost and accuracy.

\begin{table}[t]
\centering
\caption{From domain knowledge to execution: directives specify \textit{what} the user expressed; operators define \textit{how} to apply it.}
\vspace{-2mm}
\label{tab:knowledge-directives-operators}
\resizebox{\columnwidth}{!}{%
\begin{tabular}{@{}llll@{}}
\toprule
\textbf{Knowledge} & \textbf{Directive} & \textbf{Operator} & \textbf{Effect on Pipeline} \\ \midrule
Where to look & $\mathcal{S} = \{\sigma_1, \ldots, \sigma_n\}$ & $\segment(\tilde{D}, \mathcal{S}) \rightarrow D_s$ & Prunes document structure \\
What to ignore & $\mathcal{F} = \{\phi_1, \ldots, \phi_m\}$ & $\filter(\mathcal{C}, \mathcal{F}, q) \rightarrow \mathcal{C}^*$ & Discards irrelevant chunks \\
How to verify & $\mathcal{V} = \mathcal{V}^+ \cup \mathcal{V}^-$ & $\validate(\mathcal{R}, \mathcal{V}) \rightarrow A^*$ & Ranks candidate responses \\
\bottomrule
\end{tabular}%
}
\end{table}

\subsection{Optimizing Context Selection}
\label{sec:applying-domain:optimizing-context-sel}
The structural knowledge provided by the user is extracted as structural directives $\mathcal{S} = (\sigma_1, .., \sigma_n)$ by the \parser (\autoref{sec:capturing:struct}).
The \selector applies the \segment operator, using these directives to prune $\tilde{D}$ during query execution.

\noindent \textbf{Document Pruning.} The \segment operator takes $\tilde{D}$ and the structural directives $(\sigma_1, \ldots, \sigma_n)$ as input to produce the pruned document $D_s$.
Each directive $\sigma_i$ independently selects matching structural elements from $\tilde{D}$, and the operator returns their union:
\[ \segment(\tilde{D}, \mathcal{S}) = \{e \in \tilde{D} \mid 
\exists\, \sigma_i \in \mathcal{S}: \text{match}(e, \sigma_i)\} 
\rightarrow D_s \]
The \parser extracts these directives from the user's prompt as natural language descriptions (\eg $\sigma_i =$ ``EPS table'').
The \segment operator maps each directive to the concrete structural elements available in $\tilde{D}$ using embedding-based similarity matching.
Directives with no match are skipped.
If none of the directives match any element in $\tilde{D}$, the operator conservatively falls back to the full document ($D_s = D$) to avoid discarding the answer due to imprecise directives (discussed further in~\autoref{sec:fallback-manager}).

The structural pruning is \textit{model-free}, \ie the \segment operator directly maps each directive to the available tags in $\tilde{D}$ and retains only the content within matched tags.
The context reduction from the \segment operator depends on the specificity of the user directives and the document structure.
In our experiments on the SEC 10-K dataset (\autoref{sec:experimental-setup}), broad directives like ``the answer is in tables'' reduced the document from $\sim$200K to $\sim$50K tokens on average (75\% reduction). More specific directives like ``EPS table'' achieved up to 98\% reduction when matched, but yielded no reduction if no tags match.

\begin{figure}[t]
    \centering
    \includegraphics[width=0.8\columnwidth]{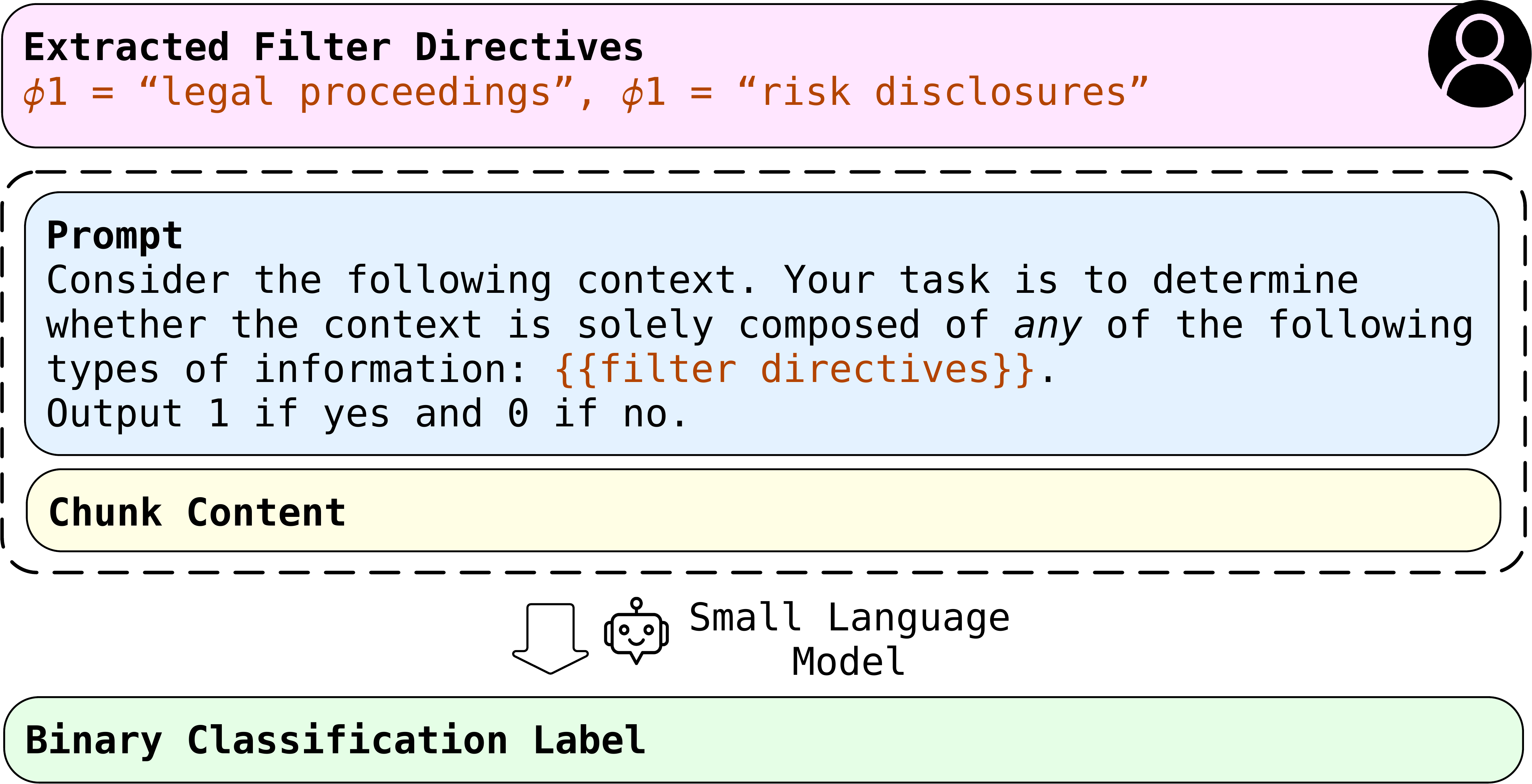}
    \caption{Applying \filter Operator}
    \vspace{-2mm}
    \label{fig:filter-operator-example}
\end{figure}

\subsection{Optimizing Inference}
\label{sec:applying-domain:optimizing-inference}
After applying the \segment operator,
the document $D_s$ retains only structurally relevant regions but may still contain semantically irrelevant content.
For example, in a financial document already filtered to tables, irrelevant tables such as the table of contents remain.
Quickly discarding such content before having an LLM generate responses enables substantial speedups, analogous to proxy-model filtering in video analytics~\cite{kang2017noscope, lu2018accelerating, kang2019blazeit}.

At a high level, the \inferenceengine achieves this through three steps: (1)~divide $D_s$ into chunks, (2)~use a lightweight model to filter out irrelevant chunks, and (3)~generate responses only for the remaining chunks using an expensive LLM.

\noindent \textbf{Chunk-Level Inference.}
Prior work has shown that LLM accuracy degrades as context length
increases~\cite{lostinthemiddle}.
Similar to RAG, \sys divides the pruned document $D_s$ into chunks $\mathcal{C} = \{C_1, \ldots, C_n\}$.
However, unlike RAG, which concatenates retrieved chunks into a single context for one inference call, \sys processes each chunk through a \textit{separate} LLM call:
\[ \{r_1, \ldots, r_n\} = \{LLM(q, C_i) \mid C_i \in \mathcal{C}\} \]
This map-reduce style QA over documents is well-established in the literature~\cite{zhou-etal-2025-llmxmapreduce, docetl,lotus, zhang2024chain}.
\sys optimizes this by leveraging domain knowledge in the \filter operator to remove irrelevant chunks during mapping (\ie before expensive inference) using the extracted filter directives.

\noindent \textbf{Applying the \filter Operator.}
The \filter knowledge provided by the user is extracted as the filter directives $\mathcal{F} = (\phi_1, \ldots, \phi_n)$---semantic descriptions of content to exclude, such as ``risk disclosures'' or ``legal disclaimers.''
To reduce inference cost, \sys evaluates each chunk against these directives using a Small Language Model (SLM) with 2--5B parameters, up to 500$\times$ smaller than frontier LLMs (>1 trillion parameters).
The reduced context size (one chunk per call) and the structurally relevant content (after \segment) simplifies the per-chunk classification task.
By simplifying the task, lightweight SLMs can filter accurately without sacrificing quality.
As shown in~\autoref{fig:filter-operator-example}, the \filter operator combines the directives into a single prompt and invokes the SLM on each chunk:
\[ \text{keep}(C_i, \mathcal{F}, q) = \text{SLM}(C_i \oplus q \oplus 
\text{serialize}(\mathcal{F})) \rightarrow \{0, 1\} \]
\[ \filter(\mathcal{C}, \mathcal{F}, q) = \{C_i \in \mathcal{C} \mid 
\text{keep}(C_i, \mathcal{F}, q) = 1\} \rightarrow \mathcal{C}^* \]

We use $\mathcal{C}^* = \filter(\mathcal{C}, \mathcal{F}, q) \subseteq \mathcal{C}$ for the filtered candidate set.

\noindent \textbf{Response Generation.}
After filtering, the retained chunks proceed to the expensive LLM for per-chunk response generation:
\[ \mathcal{R} = \text{Gen}(\mathcal{C}^*,\; q) = \{LLM(q, C_i) \mid C_i \in \mathcal{C}^*\} \]
producing candidate responses $\mathcal{R} = \{r_1, \ldots, r_m\}$, where $m = |\mathcal{C}^*|$.
Intuitively, \sys processes chunks through a \textit{model cascade}: a lightweight SLM evaluates each chunk against the filter directives, and only chunks that pass proceed to the more expensive LLM.

\subsection{Mitigating Hallucinations}
\label{sec:applying-domain:mitigating-hallucinations}
After applying the \filter operator, the retained chunks $\mathcal{C}^*$ are processed by the LLM to generate chunk-level responses $\{r_1, \ldots, r_m\}$.
The \verifier ranks these responses to select the best answer.

\noindent \textbf{The Hallucination Problem.}
The chunk-level inference pattern (\cref{sec:applying-domain:optimizing-inference}) generates multiple candidate responses, each derived from a different chunk of the document.
While the reduced context length per call improves per-chunk accuracy, the LLM can still hallucinate when a chunk contains content that is semantically related but factually distinct from the gold context (\eg a table with 2021-22 earnings when the query asks for 2022-23 earnings).
This can produce a mix of hallucinated and correct responses across chunks.

A na\"ive approach ranks these responses by model confidence, \ie the probability the LLM assigns to its generated answer.
However, LLMs are known to be confidently incorrect~\cite{chhikara2025mind}, particularly when the context contains similar but distinct information (\eg ``diluted EPS'' vs ``basic EPS'' in financial documents).
Our key insight is that users express validation knowledge in two forms: what the response \textit{should} contain (\eg ``diluted shares'', ``net revenue'') and what it should \textit{not} contain (\eg ``NOT basic EPS'', ``NOT gross revenue'').
\sys extracts and applies this validation knowledge to rank responses, with the goal of surfacing the correct answer to the top of the ranking.

\noindent \textbf{Applying the \validate Operator.}
As defined in~\autoref{sec:capturing:validate}, the validation directives are partitioned into positive constraints $\mathcal{V}^+$ and negative constraints $\mathcal{V}^-$.
Given the candidate responses, $\{r_1, \ldots, r_m\}$, \sys needs to measure how well they align with the positive and negative constraints.
To quantify this alignment, our validation score for each response rewards alignment with positive constraints and penalizes alignment with negative constraints:
\begin{align*}
\text{score}(r_i, \mathcal{V}) = \sum_{\delta \in \mathcal{V}^+} 
\text{sim}(r_i, \delta) - \sum_{\delta \in \mathcal{V}^-} 
\text{sim}(r_i, \delta)
\end{align*}
where $\text{sim}(r_i, \delta)$ denotes the cosine similarity between the embedding representations of response $r_i$ and constraint $\delta$.
Intuitively, a response mentioning ``diluted EPS'' scores high against a positive constraint $\delta_1 = \text{``diluted EPS''}$, while a response mentioning ``basic EPS'' incurs a penalty from a negative constraint $\delta_2 = \text{``basic EPS''}$.

The \validate computes the validation score for each response and returns the top-ranked responses:
\[ \validate(\mathcal{R}, \mathcal{V}) = \text{top-}k_{r \in \mathcal{R}}
\bigl(r \mid \text{score}(r, \mathcal{V})\bigr) \rightarrow A^* \]
Unlike confidence-based ranking, the \validate operator systematically leverages domain knowledge to identify the semantically correct response.
Our experiments (\autoref{sec:evaluation}) show that validation-based ranking improves accuracy by up to 8\% compared to confidence-based ranking (\autoref{tab:ranking-analysis}).
Since the scoring function is used for ranking rather than thresholding, the relative count of positive and negative constraints provided by the user does not affect correctness.
This avoids the need for manual and cumbersome threshold tuning.
For example, even with only negative constraints, the correct response incurs a smaller penalty than incorrect ones, preserving the ranking order.

\subsection{Extensions}
\label{sec:applying-domain:extensions}
While the three operators described above capture the most common prompting patterns, \parser is extensible to other knowledge types.
For example, experts may express terminology equivalence (\eg ``treat MI and myocardial infarction as equivalent'') in their prompts, which could be extracted into a \texttt{NORMALIZE} operator.
Precedence knowledge (\eg ``prioritize master agreement over the addendum'') expresses conflict resolution rules, which could be extracted into a \texttt{PRECEDENCE} operator.
Custom operators such as \texttt{NORMALIZE} and \texttt{PRECEDENCE} can be defined and implemented in \sys following the same pattern as the three operators described in this section.

\section{Operator Robustness}
\label{sec:fallback-manager}
Domain operators enhance long-context QA when the user-provided domain knowledge is accurate and well-specified.
However, in practice, operators may be imprecise, overly aggressive, or misaligned with the document content, which can degrade accuracy.
\sys addresses this through a fallback mechanism that detects imprecise operators at runtime and recovers by selectively disabling problematic operators.
This allows \sys to benefit from high-quality operators when available, while gracefully falling back to baseline performance when operators are imprecise.

\subsection{When Do Operators Degrade Accuracy?}
\label{sec:fallback:when}
We now describe how each operator can be impacted by inaccurate domain knowledge.

\noindent \textbf{\segment Operator.}
The \segment operator prunes the input document to retain only structurally relevant regions (\autoref{sec:applying-domain:optimizing-context-sel}).
This pruning removes the gold context (\ie the portion of the document containing the correct answer) when:
(1) the user specifies an incorrect structural element (\eg directing the system to look in tables when the answer appears in a text paragraph),
(2) none of the user directives match the document's structural tags, or
(3) the document parsing fails to extract the relevant structure (\eg a malformed table is not tagged).
In all cases, downstream chunk-level inference operates on a document no longer containing the answer.

\noindent \textbf{\filter Operator.}
The \filter operator discards entire chunks deemed irrelevant by the SLM-based classifier (\autoref{sec:applying-domain:optimizing-inference}).
The operator degrades accuracy when:
(1) the user's filter directives are inaccurate (\eg ``ignore financial tables'' when the answer is in a financial table), or
(2) the SLM misclassifies a relevant chunk as irrelevant, either due to model errors or because the chunk contains irrelevant portions mixed with the gold context.

\noindent \textbf{\validate Operator.}
The \validate operator ranks candidate responses using embedding similarity against positive and negative constraints (\autoref{sec:applying-domain:mitigating-hallucinations}).
Unlike the \segment and \filter operators, which affect \textit{what context} the LLM sees, \validate affects the \textit{response ordering} of candidates.
The \validate operator degrades accuracy when the constraints fail to produce meaningful score differences across candidate responses.
This occurs when:
(1) the constraints are irrelevant to the document content (\eg constraints about ``martial arts'' applied to a financial document), producing uniformly low similarity scores, or
(2) the constraints are too generic to distinguish between responses (\eg ``financial data'' when all candidate responses discuss financial content), producing uniformly high similarity scores.
In both cases, the constraints match all responses similarly, and the validation ranking is non-discriminative.

\subsection{Detecting Operator Imprecision}
\label{sec:fallback:detection}
To recover from low-quality operators, \sys must first detect that an operator has negatively impacted the query result.
We identify two detection mechanisms that leverage signals produced during normal execution, requiring no additional LLM calls.

\subsubsection{Detecting Context Loss}
\label{sec:fallback:detection:context}
To detect when \segment or \filter operators are too aggressive, \sys leverages the LLM's inherent \textit{answer refusal} capability, \ie its tendency to respond with ``answer not in context'' when the provided context is insufficient~\cite{feng2024don, cao2024learn, xu2024rejection}.
When \textit{every} chunk-level response is a refusal, the upstream operators have removed all content related to the query:

\[
\forall\, r_i \in \{r_1, \ldots, r_m\}: r_i = \text{``answer not in context''}
\]

This all-refusal signal detects complete context loss and triggers the context fallback (\autoref{sec:fallback:recovery:context}).
When at least one chunk produces a substantive response, the system proceeds to the ranking stage, where the \validate operator handles candidate selection.

\subsubsection{Detecting Ranking Degradation}
\label{sec:fallback:detection:ranking}
When \validate constraints are inaccurate, the correct response may be ranked lower than incorrect alternatives.
Unlike context loss, the LLM \textit{does} generate the correct answer, but it is not selected as the top response.

\sys uses score variance across candidate responses to detect ranking degradation:
if all responses receive similar validation scores (\ie low variance across $\{score(r_1, \mathcal{V}), \ldots, score(r_m, \mathcal{V})\}$), the constraints fail to discriminate among candidates.
When low variance is detected, \sys flags the \validate operator as non-discriminative and triggers the ranking fallback (\autoref{sec:fallback:recovery:ranking}).

\subsection{Recovery via Fallback}
\label{sec:fallback:recovery}
Once \sys detects operator imprecision, it initiates an operator-specific fallback mechanism to recover accuracy.

\subsubsection{Context Fallback}
\label{sec:fallback:recovery:context}
When context loss is detected, \sys disables the \segment and \filter operators and re-executes the query on the full document.
When the \segment operator is bypassed, the full parsed document $\tilde{D}$ is used instead of the pruned $D_s$.
When the \filter operator is bypassed, all chunks proceed to LLM inference without SLM-based filtering.
The re-execution generates a new set of candidate responses from the unfiltered document.

\subsubsection{Ranking Fallback}
\label{sec:fallback:recovery:ranking}
When non-discriminative validation is detected, \sys disables the \validate operator and falls back to confidence-based ranking.
The candidate responses $\{r_1, \ldots, r_m\}$ are re-ranked using the LLM's output probabilities:
\[
A^*_{fallback} = \text{top-}k(r_i \mid \text{conf}(r_i))
\]
where $\text{conf}(r_i)$ is the model's confidence score for response $r_i$.
In both cases, fallback ensures performance is no worse than a vanilla LLM baseline: the context fallback recovers full-document inference, and the ranking fallback reverts to confidence-based selection.
We show the effectiveness of the fallback mechanism in recovering from low quality operators in~\autoref{sec:eval-hint-quality}.

\begin{table*}[t]
\centering
\renewcommand{\arraystretch}{1.3}
\footnotesize
\caption{Representative examples of domain knowledge used in our experiments. The domain knowledge column shows the natural language instructions a domain expert provides alongside the question.}
\label{tab:exp-domain-hint-examples}
\begin{tabularx}{\linewidth}{p{0.08\linewidth} p{0.33\linewidth} X}
\toprule
\textbf{Dataset} & \textbf{Sample Question} & \textbf{Domain Knowledge} \\
\midrule
\textbf{DocFinQA} &
What was the percentage cumulative total return for the five year period ended 31-Dec-2017 of Citi common stock? &
Look in tables and the MD\&A section. Ignore legal disclaimers, table of contents, and chunks not relevant to the query. The answer should reference Citi Common Stock Cumulative Total Return from the performance graph source. Do NOT confuse with S\&P 500 return. \\
\midrule
\textbf{NovelQA} &
How long did it take for Sibyl's brother and Dorian Gray to meet after Sibyl's death? &
Sibyl's brother is James Vane. Skip chapters before Sibyl's death and chapters unrelated to James Vane or Dorian. The answer involves the time gap between Sibyl's death and James confronting Dorian. Do NOT confuse with Dorian's interactions with Lord Henry after Sibyl's death. \\
\midrule
\textbf{QASPER} &
Do they try to use byte-pair encoding representations? &
Look in the methodology and results sections. \\
\bottomrule
\end{tabularx}
\end{table*}

\section{Experiments}
\label{sec:evaluation}

\begin{table}[h!]
\centering
\small
\caption{Dataset characteristics.}
\label{tab:dataset-stats}
\resizebox{0.75\columnwidth}{!}{%
\begin{tabular}{cccccc}
\toprule
\textbf{Dataset} & \textbf{QA Domain} & \textbf{\# Queries} & \textbf{Avg.}
\\
 &  &  & \textbf{\#tokens}
\\
\midrule
DocFinQA & Finance & 100  & 193K  \\
NovelQA & Literature & 99 & 230K  \\
QASPER & Science Articles & 90 & 7K \\
\bottomrule
\end{tabular}
}
\end{table}

In this section, we evaluate \sys to empirically answer the following research questions:
\squishitemize
    \item RQ1: How does~\sys compare to baselines in end-to-end accuracy and cost across models and domains? (\autoref{sec:eval:end-to-end})
    \item RQ2: How do individual operators contribute to accuracy and cost improvements? (\autoref{sec:eval-ablation})
    \item RQ3: How robust is \sys to low-quality domain knowledge, and how effective is its fallback mechanism? (\autoref{sec:eval-hint-quality})
    \item RQ4: What is the cost overhead of \sys's operators, and how effective is validation-based ranking? (\autoref{sec:eval-overhead}, \autoref{sec:eval-ranking-analysis})
\squishend

\subsection{Experimental Setup}
\label{sec:experimental-setup}
\PP{Datasets} We evaluate \sys across three datasets from diverse domains.
~\autoref{tab:dataset-stats} summarizes the key properties of these datasets.
\squishitemize
\item DocFinQA~\cite{docfinqa}. This dataset contains full Securities and Exchange Commission (SEC) filings, leading to significantly larger document contexts. The dataset includes 801 documents with an average of 123k words per document and a total of 7,437 expert-annotated QA pairs. The output is in numerical format, allowing for a thorough evaluation of numerical reasoning over long contexts. We randomly selected 100 questions from the development set.
\item NovelQA~\cite{wangnovelqa}. This dataset contains full-length English novels averaging more than 200K tokens with different QA tasks such as single-hop QA, multi-hop QA, and summarization. We use 99 multiple-choice questions from the single-hop subset.
\item QASPER~\cite{qasper}. This dataset contains 5,049 questions over 1,585 research papers in the NLP domain, focusing on information-seeking QA tasks. Each question is grounded in full-length academic papers, with evidence spanning multiple sections, paragraphs, and figures or tables. The average document length is 7K tokens. %
We randomly selected a subset of 90 examples that are of the true/false category for reliable answer evaluation.
\squishend
\PP{Baselines} We evaluate \sys against six baselines:
\squishitemize
\item \singlepass. This baseline processes the entire document in a single LLM call and generates a single response.
We set the context length at 200K tokens, which is the context window limit for Claude Sonnet 4.5~\cite{sonnet452025}.
\item \lcrag. This baseline leverages an embedding model~\cite{bge-m3} to retrieve the top-$k$ relevant chunks, then sends the concatenated chunks as context to an LLM to generate a single response.
\item \{\singlepass, \lcrag\}+DK. These methods inject domain knowledge directly into the prompt, and call the LLM with context gathered either using the \singlepass strategy or \lcrag.
The domain knowledge is written as natural language instructions at the end of the user question, similar to how users write instructions in their prompts today.
\item \multiagent~\cite{zhang2024chain}. Chain-of-Agents (CoA) processes individual chunks of long documents via worker LLMs with sequential information sharing across worker agents.
It then aggregates results via a manager LLM.
We adapt COA for pinpointed QA by using parallel workers, since sequential evidence aggregation provides no benefit when answers are localized to single chunks.
Workers process the full document without retrieval.
\item \halonodk. This baseline uses the \sys pipeline without enabling the domain operators.
The top-$k$ chunks are selected via embedding similarity and processed via individual LLM calls, and the responses are ranked using confidence score (for local models) and in the selection order (for closed models).
\squishend

For all baselines that require chunking and top-$k$ chunk selection, we perform table/layout-aware chunking, use a chunk size of $\sim$4000 tokens, and set $k = 10$.
Baselines that process chunks via independent LLM calls---\multiagent, \halonodk, and \sys---leverage batched inference to reduce per-query latency and cost.
All chunk-level calls within a single query are issued in parallel, and the reported costs for these baselines reflect this batched execution.

\PP{Domain Knowledge Sources}
For each dataset, we construct domain knowledge prompts as would be provided by domain experts alongside their question.
These prompts are constructed through manual inspection of the dataset queries and source documents, supplemented by dataset-specific metadata where available.
For DocFinQA, we use evidence annotations from the parent dataset FinQA~\cite{chen2021finqa} to determine whether the answer resides in a table, text, or both, and construct structural directives accordingly (\eg ``look in tables and the MD\&A section'').
We derive filter directives from recurring irrelevant content patterns in SEC 10-K filings (\eg legal disclaimers, tables of contents).
Validation directives are constructed by inspecting the evidence text associated with each query to identify entities that co-occur with the correct answer (positive directives, \eg ``diluted computations'') and nearby entities that could be confused with it (negative directives, \eg ``basic computations'').
For NovelQA, we use chapter-level metadata and character annotations to construct directives that guide the system to relevant portions of the novel (\eg ``skip chapters before Sibyl's death'').
For QASPER, we use section-level structure common to NLP papers (\eg ``look in the methodology and results sections'').
\autoref{tab:exp-domain-hint-examples} provides representative examples from each dataset.
On average, the \parser extracts 10, 6, and 2 directives per query for DocFinQA, NovelQA, and QASPER, respectively.

\PP{Implementation Details} We use both open- and closed-source LLMs for our experiments.
For the closed-source model, we use Claude Sonnet 4.5~\cite{sonnet452025} (released September, 2025) via the Amazon AWS Bedrock Platform API~\footnote{https://aws.amazon.com/bedrock/}.
For the open-source model, we use the \texttt{Qwen3-30B-A3B-Instruct} long-context model~\cite{qwen3technicalreport} (released July, 2025) hosted locally with the vLLM inference framework~\cite{kwon2023efficient}.
We use \texttt{Qwen3-4B-Instruct} hosted locally on vLLM as the lightweight filter model for our \filter operator.
For selecting the top-$k$ relevant chunks, we use the bge-m3 embedding model~\cite{bge-m3}.
We use the \texttt{sentence-transformers} library for embedding computations.

\PP{Metrics} We use Mean Reciprocal Rank (MRR@k) to measure accuracy, a standard metric for evaluating ranked response lists~\cite{khattab2020colbert}.
MRR@k measures how often the correct answer appears within the top-$k$ ranked results; higher values at low $k$ indicate that users need to inspect fewer responses.
For DocFinQA, we use relative error with 5\% tolerance since the responses are numerical answers.
For NovelQA and QASPER, we perform exact matching against the ground truth.
For cost, we report the average dollar cost per query based on AWS Bedrock pricing\footnote{https://aws.amazon.com/bedrock/pricing/}, computed from total input and output tokens across all LLM and embedding model calls.

\subsection{End-to-end Results}
\label{sec:eval:end-to-end}
\begin{figure}[t]
    \centering
    \includegraphics[width=0.9\columnwidth]{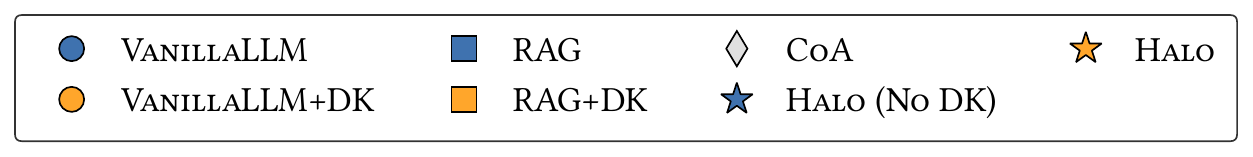}
    \includegraphics[width=0.9\columnwidth]{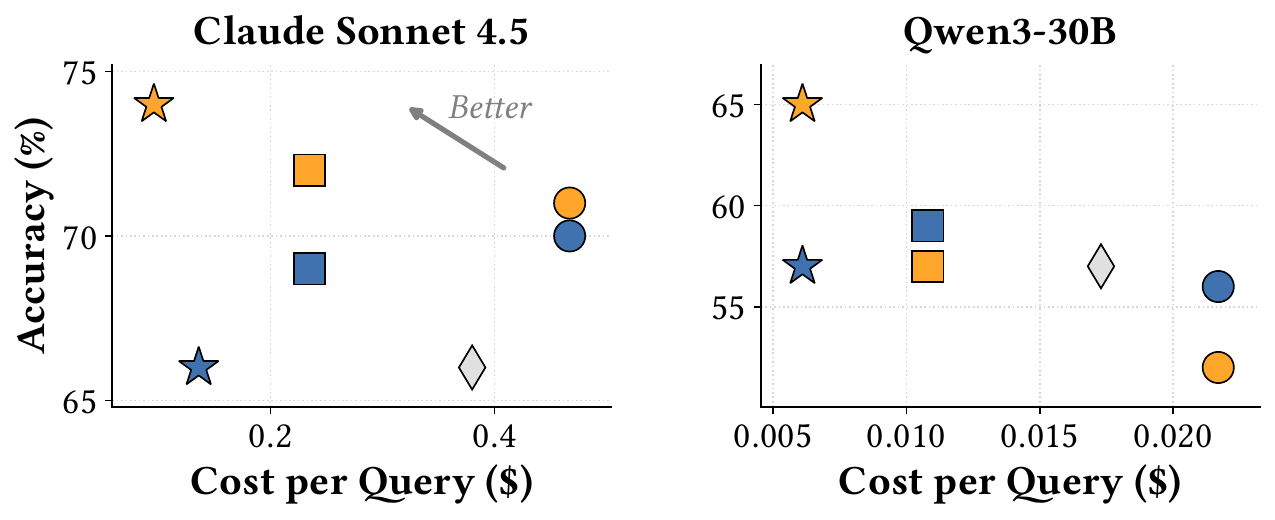}
    \caption{End-to-end accuracy (MRR@1) vs.\ cost per query on DocFinQA (100 queries). \sys achieves the highest accuracy at the lowest cost across both LLMs (top-left is best).
    }
    \label{fig:endtoend-results}
\end{figure}

We first explore how \sys uses domain knowledge to improve query accuracy and cost using the DocFinQA dataset.
We use two LLMs: Claude Sonnet 4.5~\cite{sonnet452025} and Qwen3-30B~\cite{qwen3technicalreport}.
\autoref{fig:endtoend-results} shows the accuracy (MRR@1) and cost per query for all methods.

By capturing and applying domain knowledge in query processing, \sys improves accuracy over \singlepass by 4\% and 9\% using Claude and Qwen, respectively.
\sys also reduces cost by up to 4.8$\times$ compared to \singlepass.
Compared to \lcrag, it improves accuracy by 9\% and cost by up to 1.8$\times$.
\singlepass and \lcrag process the document without any domain knowledge, incurring high cost and suffering from accuracy degradation over long contexts.

Even when injecting the domain knowledge into the baselines' prompts (\singlepassdk and \lcragdk), \sys achieves up to 4.8$\times$ lower cost and up to 13\% higher accuracy.
Further, injecting the domain knowledge can sometimes \emph{hurt} the baselines accuracy: with Qwen, \singlepassdk dropped by 4\% compared to \singlepass, and \lcragdk dropped by 2\% compared to \lcrag.
This is consistent with prior work that shows weaker models struggle to follow complex instructions over long contexts~\cite{he2024can,yen2025helmet}.
This demonstrates that systematically applying the domain knowledge is necessary for reducing accuracy degradation.
Also of note: both +DK methods incur the same cost as their non-domain knowledge counterparts since the model still processes the entire document.

Compared to \multiagent, \sys improves accuracy by 8\% on Claude and reduces cost by 4$\times$.
\multiagent achieves the \textit{lowest} accuracy on Claude (66\%), underperforming even \singlepass (70\%), despite decomposing the document into chunks.
We observe that while the candidate responses do contain the correct answer, the manager LLM often fails to identify it from the large pool of candidates.
\sys's \validate operator addresses this gap by leveraging domain directives to identify the correct response from the candidate pool.
\multiagent also incurs significant costs even with chunk-level batch processing enabled due to processing the \textit{entire document} sequentially in chunks (as opposed to \singlepass that truncates the document).

\PP{Robustness across models}
\sys's improvements on cost and accuracy hold regardless of the underlying LLM.
The accuracy gap between \sys and baselines is larger with Qwen than with Claude, suggesting that weaker models benefit more from systematic domain knowledge application, as they are less capable of implicitly leveraging domain context on their own.
\sys with Qwen (65\%) approaches \singlepass on Claude (70\%) at 78$\times$ lower cost, demonstrating that \sys can enable a smaller, more cost-efficient model to approach the accuracy of a frontier model used naively.

\takeaway{\sys consistently achieves the best accuracy--cost trade-off across all baselines. Systematically applying domain knowledge is more effective and robust than prompt injection or engineering, enabling a lightweight open-source model to approach frontier accuracy at 60--80$\times$ lower cost.}

\begin{table}[t]
\centering
\small
\caption{Domain generalization: accuracy (\%) and cost (\$/query) on NovelQA (99 queries, $>$200K tokens) and QASPER (90 queries, 7K tokens) using Qwen. Values in parentheses show the change when adding prompt-based domain knowledge (+DK).}
\label{tab:domain-gen-results}
\resizebox{0.6\columnwidth}{!}{%
\begin{tabular}{lcccc}
\toprule
& \multicolumn{2}{c}{\textbf{NovelQA}} & \multicolumn{2}{c}{\textbf{QASPER}} \\
\cmidrule(lr){2-3} \cmidrule(lr){4-5}
\textbf{Method} & \textbf{Acc.} & \textbf{Cost} & \textbf{Acc.} & \textbf{Cost} \\
\midrule
\singlepass       & 80\,\small{(\textminus1)} & 0.029 & 64 & 0.0012 \\
\lcrag            & 78\,\small{(\textminus1)} & 0.029 & 64 & 0.0012 \\
\multiagent       & 79 & 0.028 & 64 & 0.0011 \\
\halonodk         & 79 & 0.023 & 64 & 0.0011 \\
\sys              & \textbf{83} & \textbf{0.022} & 64 & \textbf{0.0007} \\
\bottomrule
\end{tabular}%
}
\end{table}

\PP{Performance on other domains}
We also evaluate \sys on NovelQA (literature, $>$200K tokens) and QASPER (scientific articles, 7K tokens) using the Qwen model.
The results are in~\autoref{tab:domain-gen-results}.
On NovelQA, \sys outperforms all baselines by 4\% in accuracy.
The gains are smaller than on DocFinQA because narrative text offers less exploitable structure for the \segment operator, and entity references are distributed throughout the text, limiting \filter effectiveness.
However, \sys's \validate operator remains effective at detecting hallucinations in narrative-style responses, where LLMs tend to fabricate elaborate but incorrect evidence.
On QASPER, all methods achieve similar accuracy due to the short context length (7K tokens).
\sys still provides cost benefits through \segment and \filter, which avoid irrelevant sections and chunks without sacrificing accuracy.

\takeaway{\sys's benefits are most pronounced on long, structured documents. On narrative text, \validate remains effective at detecting hallucinations even when structural operators offer limited gains.}

\begin{table}[t]
\centering
\small
\caption{Contribution of operators on DocFinQA (100 queries, Claude): accuracy (\%) and cost (\$/query) when each operator is applied independently to the \halonodk pipeline.}
\label{tab:operator-ablation}
\resizebox{0.6\columnwidth}{!}{%
\begin{tabular}{lcc}
\toprule
\textbf{Configuration} & \textbf{Acc.\ (\%)} & \textbf{Cost (\$/q)} \\
\midrule
\halonodk              & 66 & 0.14 \\
\quad + \segment       & 69 & 0.12 \\
\quad + \filter        & 65 & 0.11 \\
\quad + \validate      & 71 & 0.14 \\
\midrule
Full \sys              & 74 & 0.10 \\
\bottomrule
\end{tabular}%
}
\end{table}

\subsection{Contribution of Individual Operators}
\label{sec:eval-ablation}
We now independently assess the impact of each domain operator using DocFinQA with Claude.
\halonodk runs \sys's full multi-stage pipeline---chunking, parallel inference, and confidence-based ranking---without any domain operators, serving as the baseline for measuring operator-level gains.
\autoref{tab:operator-ablation} shows the accuracy and cost per query for each configuration.

\PP{\segment}
Adding the \segment operator improves accuracy by 4\% while reducing cost by 14\%.
By pruning the document to retain only structurally relevant regions before chunking, the \segment operator produces context-rich chunks, reducing hallucinations in individual chunk-level responses.
This also reduces token cost in cases where the total chunks are already below the top-$k$ due to aggressive pruning at the \segment level.

\PP{\filter}
The \filter operator reduces cost by 21\% with a 1\% accuracy drop.
Unlike \segment which operates before top-$k$ selection, \filter removes chunks directly from the $k$ selected chunks before expensive LLM calls, yielding larger per-query cost reduction.
The slight accuracy drop occurs because the SLM-based classifier may occasionally discard a chunk containing relevant information.

\filter may discard relevant responses if the SLM-based filtering is inaccurate.
To mitigate this, \sys applies \filter conservatively, only removing chunks that are highly likely to be irrelevant.

\PP{\validate}
The \validate operator improves accuracy by 5\% with no cost impact.
Unlike \segment and \filter, which reduce the context seen by the LLM, \validate re-ranks candidate responses using domain-specific constraints~(\autoref{sec:applying-domain:mitigating-hallucinations}) to effectively detect and demote hallucinated responses.
By rewarding responses aligned with positive constraints (\eg ``diluted EPS'') and penalizing those matching negative constraints (\eg ``basic EPS''), \validate surfaces correct answers that confidence-based ranking misses.
The cost impact is negligible since ranking only requires computing embedding similarities over the already-generated responses~(\cref{sec:applying-domain:mitigating-hallucinations}).

\PP{Combined effect}
Full \sys applies all three operators, achieving 74\% accuracy at \$0.10/query---an 8\% accuracy gain and 29\% cost reduction over \halonodk.
Notably, the accuracy gains from \validate compound when combined with richer inputs from \segment and \filter: \segment improves per-chunk response quality by providing cleaner context, which in turn enables \validate to more effectively rank the responses using domain constraints.
The cost reductions from \segment and \filter also compound, as they operate at different granularities---document-level structural pruning and chunk-level semantic filtering, respectively.
\takeaway{Each operator contributes distinct benefits: \segment focuses the context, \filter reduces the cost via context pruning, while \validate improves accuracy through domain-aware ranking. Their effects compound in the full system.}

\begin{table}[t]
\centering
\caption{Robustness to low-quality operators on DocFinQA (Qwen): accuracy (\%) and cost (\$/query) when \sys is given incorrect \segment elements and irrelevant \validate constraints, with and without the fallback manager.}
\vspace{-2mm}
\label{tab:operator-robustness}
\resizebox{0.6\columnwidth}{!}{%
\begin{tabular}{l cc cc}
\toprule
& \multicolumn{2}{c}{\textbf{w/o Fallback}} & \multicolumn{2}{c}{\textbf{w/ Fallback}} \\
\cmidrule(lr){2-3} \cmidrule(lr){4-5}
\textbf{Operator} & \textbf{Acc.} & \textbf{Cost} & \textbf{Acc.} & \textbf{Cost}\\
\midrule
\segment   & 14\% & 0.007 & 55\% & 0.017 \\
\validate  & 50\% & 0.006 & 57\% & 0.006 \\
\bottomrule
\end{tabular}%
}
\end{table}

\subsection{Robustness to Low-Quality Operators}
\label{sec:eval-hint-quality}
We now evaluate \sys's ability to detect low-quality operators and use its fallback manager~(\autoref{sec:fallback-manager}) to prevent accuracy degradation.
We consider two types of low-quality operators:
For \segment, we provide incorrect structural elements (\eg \texttt{text} for table-answered questions), which causes the operator to prune the gold context.
For \validate, we provide constraints unrelated to the document content (\eg constraints about science articles applied to a financial document), making the operator non-discriminative.
We evaluate \sys on the 100 queries from the DocFinQA dataset using Qwen.

\autoref{tab:operator-robustness} shows the accuracy and cost per query of \sys without and with its fallback manager.
\sys effectively handles low-quality \segment operators in~80\% of the cases via the context-loss detection mechanism~(\autoref{sec:fallback:detection:context}), which leads to an accuracy recovery from 14\% to 55\%.
In the remaining 20\%, the pruned document still contains content semantically related to the query (\eg a text paragraph mentioning EPS when the answer resides in a table), causing the LLM to generate a confident but incorrect response rather than refusing, and thus preventing the all-refusal signal from triggering.

Leveraging model abstention and confidence scores can effectively detect and mitigate the negative impact of filtering out relevant context.
This recovery process incurs a 1.7$\times$ increase in cost due to necessary re-execution on the full document~(\autoref{sec:fallback:recovery:context}).

\sys shows similar robustness to low-quality \validate operators.
When irrelevant constraints are provided, the validation scoring function~(\autoref{sec:applying-domain:mitigating-hallucinations}) produces near-zero similarity with all candidate responses.
\sys detects this~(\autoref{sec:fallback:detection:ranking}) and disables the \validate operator, falling back to confidence-based ranking, which is already available from inference.
The recovered accuracy (57\%) matches \halonodk, confirming that the fallback reverts to the no-domain knowledge baseline when \validate constraints are non-discriminative.
\takeaway{\sys recovers gracefully from imprecise \segment and \validate operators. The fallback manager ensures \sys performs no worse in accuracy than the pipeline without domain operators.}

\begin{table}[t]\centering\small
\caption{Cost Analysis for \sys: Cost breakdown of \sys's individual stages and operations as a fraction of the overall query cost.}
\label{tab:cost-analysis}
\resizebox{\columnwidth}{!}{%
\begin{tabular}{lcccc}
\midrule
\textbf{Stage} & \textbf{Operator/Step} & \textbf{Model} & \textbf{Cost/query (\$)} & \textbf{\% of total} \\
\midrule
\multicolumn{5}{c}{\textbf{One-time (amortized)}} \\
\midrule
Pre-processing & - & Gemini-2.0-Flash & 0.01142 & - \\
\midrule
\multicolumn{5}{c}{\textbf{Per Query}} \\
\midrule
\parser     & - & Sonnet-4.5 & 0.001 & 1.04\% \\
\cmidrule(lr){2-5}
\multirow{2}{*}{\selector} & \segment     & bge-m3 & 0.00002 & 0.02\% \\
\cmidrule(lr){2-5}
\multirow{2}{*}{\inferenceengine} & \filter     & Qwen-4b & 0.00093 & 0.96\% \\
                                   & Expensive LLM  & Sonnet-4.5 & 0.09454 & 97.96\% \\
\cmidrule(lr){2-5}
\verifier &  \validate   & bge-m3 & 0.00001 & 0.01\% \\
\bottomrule
\end{tabular}%
}
\end{table}

\begin{table}[h!]
\centering
\small
\caption{Ranking quality at different MRR cutoffs on DocFinQA (100 queries, Claude).}
\label{tab:ranking-analysis}
\resizebox{0.6\columnwidth}{!}{%
\begin{tabular}{lccc}
\toprule
\textbf{Method} & \textbf{MRR@1} & \textbf{MRR@3} & \textbf{MRR@5} \\
\midrule
\halonodk  & 66 & 71 & 72 \\
\sys  & 74 & 76 & 76 \\
\bottomrule
\end{tabular}%
}
\vspace{-1em}
\end{table}
\subsection{Ablation Studies}

\subsubsection{Cost Breakdown}
\label{sec:eval-overhead}
We now quantify the cost of \sys's stages.
We report the breakdown of per-query cost on DocFinQA in~\autoref{tab:cost-analysis}.
We report costs using API-based pricing for all components.

The dominant cost is the expensive LLM inference (97.96\% of total). %
All other \sys operators contribute only 2.04\% combined to the total cost.
The \parser, which uses an LLM to extract domain operators from the user domain instructions, contributes only 1\% of the total cost, since user instructions rarely exceed 500 tokens compared to the $>$200K document sizes.
The \segment and \validate operators rely on lightweight embedding models for similarity over short input snippets (structural tags/responses), only taking up 0.02\% and 0.01\% of the total cost, respectively.
The \filter operator invokes a 4B-parameter SLM per candidate chunk, contributing 1\% to the total cost due to the 75$\times$ cheaper SLM cost. %
The one-time pre-processing cost covers PDF parsing and structural tag extraction using a cost-effective model~\cite{filimonov2025gemini}, and is amortized across all queries on the same document.
With a total operator cost of just \$0.003 per query, the domain operators reduce the per-query expensive LLM cost from \$0.14 (\halonodk) to \$0.095 --- a 14$\times$ return despite the added overheads.

\subsubsection{Ranking Analysis}
\label{sec:eval-ranking-analysis}
We now show \sys's ability to rank candidates as $k$ varies.
\autoref{tab:ranking-analysis} compares \sys and \halonodk for varying MRR@$k$.

\sys improves MRR consistently across all cutoffs (+5–8\% over \halonodk), confirming that the correct answers are reliably surfaced in the top positions due to top-$k$ selection with richer chunks (w/ \segment) and better re-ranking (w/ \validate).
We also observed that applying \validate alone to \multiagent improves MRR@1 by 6\%, thus demonstrating the re-ranking remains effective even over larger, noisier candidate pools.

\section{Related Work}
\label{sec:related-work}

\PP{LLMs and Data Management}
There have been extensive research efforts into leveraging LLMs to improve data management, and to incorporate data management principles for improving LLM-based pipelines.
Evaporate~\cite{evaporate} generates structured views from heterogeneous data lakes using LLMs.
ZenDB~\cite{zendb} extracts semantic hierarchical structures from semi-structured documents and utilizes LLMs to enable efficient SQL querying over these documents.
In contrast, while \sys may leverage existing structure in the document to optimize long-context QA (\eg using the \segment operator), it does not focus on the structured data extraction task.

Other works explore optimizing complex LLM-based pipelines using data management techniques.
LOTUS~\cite{lotus} introduces a semantic operator model for AI-based workloads and proposes several optimizations for each operator that reduce the cost while providing accuracy guarantees.
Palimpzest~\cite{palimpzest} and DocETL~\cite{docetl} provide declarative interfaces to define complex AI processing pipelines.
While Palimpzest uses cost-based optimization to enhance diverse AI applications, DocETL uses agent-based rewriting to optimize multi-step document processing pipelines.
\sys differs from these systems in that it automatically extracts domain knowledge from user prompts, applies it across its multi-stage pipeline, and detects accuracy degradation via its fallback mechanism.

\PP{Long-context QA}
While LLMs feature expanded context windows~\cite{googleGemini2024, qwen2.5-turbo}, effectively applying them to QA tasks over long documents poses several challenges.
Models often struggle to retain information across extensive contexts~\cite{an2025why, lostinthemiddle, kuratov2024babilong}.
RAG is a popular paradigm for QA over large information corpora~\cite{ragoriginal}, wherein relevant segments are retrieved through embedding-based similarity search and passed to an LLM for generation.
While early RAG systems performed retrieval over smaller chunks, recent advances in context window sizes have improved embedding models~\cite{bge-m3}, enabling enhanced RAG systems tailored for longer inputs.
To overcome the limitations of vanilla LLMs and RAG, multi-agent pipelines have been proposed for long-context QA~\cite{zhang2024chain} that split the document into individual chunks, process the individual chunks using multiple agent LLMs, and aggregate the responses using a manager LLM.
\sys leverages user-specified domain knowledge to improve various stages (\eg chunk selection, answer validation) within the multi-agent QA pipeline.

\PP{Domain knowledge in query optimization}
Commercial database systems have historically enabled users to express domain knowledge through user-defined hints~\cite{oracleHints, microsoftHints2025, pghintplan}.
Traditionally, these hints are designed for database administrators and domain experts to influence the query optimizer's selection of execution plans~\cite{microsoftHints2025, bruno2009power}.
More recently, this concept has been expanded to optimize video analytics pipelines that involve machine learning models.
VIVA~\cite{romero2022optimizing} introduces relational hints (\texttt{CAN REPLACE}, \texttt{CAN FILTER}) that allow users to declaratively specify model relationships and automatically selects the best plan that meets user-specified accuracy requirements.
Similarly, ClueVQS~\cite{chao2024optimizing} proposes user-defined \textit{Clues} to incorporate domain-specific knowledge into video analytics applications.
While \sys draws inspiration from these approaches, it automatically extracts domain knowledge from users' natural language prompts and applies it across multiple pipeline stages—requiring no change to how users express their knowledge.

\PP{Domain Knowledge in LLMs}
Prompting techniques like chain-of-thought have been shown to elicit more complex reasoning from LLMs~\cite{wei2022chain, kojima2022large}.
However, these techniques are sensitive to phrasing, and users often struggle to effectively articulate complex requirements or domain constraints for LLMs~\cite{ma2024you}.
This can result in multiple iterations of prompt engineering.
\sys removes the need for prompt engineering by automatically extracting domain knowledge from user prompts and applying it at the appropriate stage of its multi-stage pipeline.

\section{Conclusion}
Despite rapidly expanding context windows in modern LLMs, long-context question answering remains a significant challenge due to accuracy degradation over long contexts and high processing costs.
This paper presented \sys, a framework that bridges the gap between the domain knowledge experts naturally express in their prompts and how long-context QA systems utilize it.
By extracting this knowledge into structured directives and applying them via dedicated operators---\segment, \filter, and \validate---across distinct pipeline stages, \sys moves beyond treating domain knowledge as passive prompt tokens that the LLM may or may not follow.
Across finance, literature, and scientific domains, we demonstrate that \sys outperforms both vanilla and prompt-engineered baselines, achieving up to 13\% higher accuracy and 4.8$\times$ lower cost.
Notably, \sys enables a lightweight open-source model to approach frontier LLM accuracy at 78$\times$ lower cost, showing that systematic application of domain knowledge can compensate for model scale.

\bibliographystyle{ACM-Reference-Format}
\bibliography{main}

\appendix

\end{document}